\documentclass{appolb}
\usepackage{epsfig}
\begin{document}
\title{Statistical hadronization with resonances%
\thanks{Presented at the XLIV Crakow School of Theoretical Physics}%
}
\author{Giorgio Torrieri
\address{Department of Physics, University of Arizona, Tucson, Arizona, 85721, USA,\\ and\\
Department of Physics, McGill University, Montreal, QC H3A-2T8, Canada}
\and
Jean Letessier $^{a,b}$ , Johann Rafelski $^{a,b}$  and Steve Steinke $^a$
\address{  $^a$ Department of Physics, University of Arizona, Tucson, Arizona, 85721, USA,\\ and\\ 
 $^b$ Laboratoire de Physique Th\'eorique et Hautes Energies\\
Universit\'e Paris 7, 2 place Jussieu, F--75251 Cedex 05, France}
}
\maketitle

\begin{abstract}
We introduce the equilibrium and non-equilibrium statistical
 hadronization picture of particle production in 
ultra-relativistic heavy ion collisions. 
We describe the related physical reaction scenarios, 
and show how  these can lead to  quark pair yield non-equilibrium.
Using the SHARE1.2  program suite we
quantitatively model particle yields and ratios for RHIC-130 run.
 We study how experimental particle ratios can differentiate
between   model scenarios, and discuss
in depth the importance of hadronic resonances in understanding of
hadron production processes.
\end{abstract}

\section{Introduction}

In strong interaction reaction  processes particle production is abundant.
In the Fermi-Hagedorn  statistical model the non-perturbative description of 
particle yields is based on the assumption that particle production 
is governed solely by the size of the accessible phase space. 
For an introduction into the literature and history of the
statistical hadronization model
we refer the reader to Ref. \cite{KMR03}. Further  recent developments are 
discussed in  \cite{wbwfhag,Letessier:gp}. We focus in this report on 
current developments related to RHIC experimental program, involving
the measurement of
hadron yields, obtained by integrating the produced
stable particle spectra. 

The statistical hadronization picture is necessarily an integral part
of the modeling of this``soft'' experimental hadron  production  data.
This is both to provide a bottom line to eventual  microscopic
production mechanisms signaling new physics, and to search for abrupt 
changes   associated with a phase transformation in which hadrons dissolve
into a phase of quarks and gluons.
  
Hagedorn \cite{hagedorn}  was first to
recognize the importance of hadronic resonances in statistical particle production.
To quantitatively describe hadron final state yields, contributions from resonance
decays must be 
taken into account.    As it turns out, the effect of these resonance decays on particle
phase space is crucial:
As Hagedorn  has shown, it can be 
expected that the density density of states of strongly interacting 
resonances  increases exponentially with energy, 
\begin{equation}
\rho(E) \sim e^{\beta_{\rm H} E}
\label{rhoE}
\end{equation}
This  has  a profound
implications for the behavior of matter 
at high temperature. It leads to a phase singularity at 
the temperature (``Hagedorn Temperature'', 
$T_{\rm H}=\beta_{\rm H}^{-1} \approx 160$ MeV) where the 
canonical partition function diverges:
\begin{equation}
Z = \sum_{i} \rho(E_i) e^{- E_i/T}\left.\right|_{T\to T_{\rm H}}  \to \infty .
\label{hagZ}
\end{equation}
Hagedorn referred to this as the boiling point of hadronic matter.
With advancement of the quark picture of matter he recognized 
this singularity as being the  deconfinement 
transition, where 
hadrons cease to be fundamental degrees of freedom, 
and quarks are able to propagate freely throughout the system.  

Currently, an intense experimental effort (covering an 
energy range from 2 to 200 GeV) is underway to create,
and  identify, a relatively large volume of deconfined
quark--gluon matter, the quark--gluon plasma (QGP), and to explore its
properties. In the following section we describe in turn within 
the statistical hadronization model:
particle and quark chemistry in subsection \ref{partdis},  
resonance decay chains in  subsection \ref{resdec}, 
and we introduce   the  finite resonance widths in  subsection \ref{widthdis}.  
The reader is referred to Refs.~\cite{schne}
-- \cite{zakopane} for many further details.
In subsection \ref{alternatives} we discuss in some detail
how it is possible that chemical non-equilibrium occurs in a
strongly interacting system. We than turn to the more phenomenological aspects, 
and introduce the practical aspects of the statistical model in 
section \ref{model} where we also develop strategies for detecting
chemical nonequilibrium.  In section \ref{SHARE} we  discus 
the fitting method and proceed to present some RHIC related results 
obtained with the SHARE suite of programs. 

\section{The statistical particle production model}\label{particle}
\subsection{Particle chemistry and yields in the statistical model}
\label{partdis}
In the grand canonical description the expected baryon and meson yields 
 are given by the Fermi-Dirac or, respectively, Bose-Einstein
distribution functions:
\begin{eqnarray}
\label{nmdef}
n(m_i,g_i;T,\Upsilon_i)\equiv n_i &=& g_i \int {d^3p \over (2 \pi)^3}
{1 \over \Upsilon^{-1}_i
\exp(\sqrt{p^2+m^2_i} / T ) \pm 1} ,\\
&=& \frac{g_i}{2 \pi^2}  \sum_{n=1}^{\infty} (\mp)^{n-1}
\Upsilon_i^{n} \, \frac{T \, m_i^2}{n} \,
K_2 \left( \frac{n m_i}{T} \right).
\label{ni}
\end{eqnarray}
In  the upper signs refer to fermions and the lower signs to 
bosons, respectively. The second 
form, Eq.\,(\ref{ni}), expresses the momentum integrals in
terms of the modified Bessel function $K_2$. This form is practical in
the numerical calculations. The 
series expansion (sum over $n$) converges when $\Upsilon_i e^{-m_i/T}<1$,
below the Bose-Einstein condensation limit.
Consideration of this condition is required only for the pion yield
in the range of parameters of interest.  
Here, the index `$ i$' labels different particle species, including 
hadrons which are stable under 
strong interactions (such as pions, kaons, nucleons or hyperons) and hadron 
which are unstable ($\rho$ mesons, $\Delta(1232)$, etc.).
$m_i$ is the particle mass, while the 
quantity $g_i=(2J_i+1)$ is  the spin degeneracy factor; we will 
distinguish all isospin sub-states separately.

$\Upsilon_i$ is the fugacity factor. Since  particle anti-particle pairs ($i \overline{i}$) 
can be produced  by particles  without conserved `charges' (baryon number, strangeness, etc.),
\begin{equation}
\label{detbalance}
i + \overline{i} \rightleftharpoons n\times \gamma,g, \pi, ... \mathrm{(n-objects\;\;with\;\;no \;\;conserved\;\;charges)}
\end{equation}
a scenario where statistical production arises as a limit of kinetic evolution requires that  
the  chemical potentials of particles and antiparticles associated with the 
conserved charge be opposite. Thus the  fugacity 
$\Upsilon_i$  comprises the  well known factor   
$\lambda^{\pm1}=e^{\pm\mu/T}$, where 
$\mu$ is the chemical potential
associated with a conserved quantum number.

For a dynamically  evolving system  yield of particles 
cannot in general be in chemical equilibrium. This is  accommodated by
introducing a phase space occupancy factor $\gamma$:
\begin{equation}
\label{gamma}
\Upsilon_j  =  \gamma_j \lambda_j,\qquad  \Upsilon_{\overline{j}}= \gamma_j  \lambda_j^{-1}
\end{equation}
$\gamma_j=1$ corresponds to chemical equilibrium.   
$\gamma_j<1$ means that quantum number $j$ is under-saturated,
usually because   time is needed to build it up. An isolated system 
  kept long enough  ``in a box'' would generate more
of  $j \overline{j}$ particle-antiparticle pairs at the 
expense of the internal energy.
If this equilibrated box  undergoes isentropic and rapid expansion 
($VT^3=\mbox{Const.}$ for massless particles),  
$\gamma_j>1$ will generally result as the particle abundances
do not have time to re-equilibrate \cite{cool}.  Given time in 
the ``large box'', pairs would annihilate and the system would 
again relax to  an equilibrium  state.   
In subsection \ref{alternatives} we will explore these scenarios
more thoroughly, and in section \ref{model} we will argue 
that equilibrium assumptions can and should  be tested against experimental data.

Each hadron can be a carrier of several  conserved charges  contained in 
its  valance quarks. The most general fugacity $\Upsilon_j$ for hadron   $j$ is  
given by a suitable product of $\lambda$  and $\gamma$ factors:
\begin{equation}
\Upsilon_j=\lambda_{I^j_3} \left(\lambda_q \gamma_q\right)^{N^j_q}
\left(\lambda_s \gamma_s\right)^{N^j_s} 
\left(\lambda_{c} \gamma_{c}\right)^{N^j_{c}}
\left(\lambda_{\bar q} \gamma_{\bar q}\right)^{N^j_{\bar q}}
\left(\lambda_{\bar s} \gamma_{\bar s}\right)^{N^j_{\bar s}}
\left(\lambda_{\bar c} \gamma_{\bar c}\right)^{N^j_{\bar c}}, 
\label{upsilons}
\end{equation}
where
\begin{equation}
\lambda_{q}=\lambda^{-1}_{\bar q},\qquad
\lambda_{s}=\lambda^{-1}_{\bar s},\qquad 
\lambda_{c}=\lambda^{-1}_{\bar c},\qquad 
\label{lambdas}
\end{equation}
and
\begin{equation}
\gamma_{q}=\gamma_{\bar q},\qquad
\gamma_{s}=\gamma_{\bar s},\qquad
\gamma_{c}=\gamma_{\bar c}.
\label{gammas}
\end{equation}
Here, $N^j_q$, $N^j_s$ and $N^j_c$ are the numbers of light $(u,d)$, strange
$(s)$  and charm $(c)$ valance quarks in the $j$th hadron, and $N^j_{\bar q}$, 
$N^j_{\bar s}$  and $N^j_{\bar c}$ are the numbers of the corresponding valance 
antiquarks in the same hadron.

The quark and hadron chemical potentials as used in {\it e.g.}  Refs.~\cite{PBM99,PBMRHIC,wbwf} are   related by simple algebraic 
conditions arising from the valance quark content and definitions 
of the baryon and hyperon charges of particles given in the 
historical context. Considering the case of chemical 
equilibrium limit  $\gamma_{q}=\gamma_{s}=1$  and in absence of charm
($N^i_{c}=N^i_{\bar c}=0$):
\begin{eqnarray}
\Upsilon^{\rm \,\,eq}_i=
\exp \left( \frac{B_i \mu_B + S_i \mu_S + I^{\,i}_3 \mu_{I_3}}{T}\right),
\label{upsieq}
\end{eqnarray}
where $B_i$, $S_i$, and $I^{\,i}_3$ are the baryon number,
strangeness, and the third component of the isospin of the $i$th
particle, and $\mu$'s are the corresponding chemical potentials.  In
this case, the two formulations are related by equations:
\begin{equation}
\lambda^{\rm eq}_q = e^{\mu_B/3T}, \,\,\,\,\, 
\label{lameq}
\end{equation}
\begin{equation}
\lambda^{\rm eq}_S = e^{(-3 \mu_s + \mu_B)/3T}, \,\,\,\,\, 
\label{gameq}
\end{equation}
%
\begin{equation}
\lambda^{\rm eq}_{I^i_3} = \lambda_{I^i_3} = 
e^{I^{\,i}_3 \mu_{I_3}/T} .
\label{lami3eq}
\end{equation}
We note that strangeness and hyperon number $S$ of a particle 
are by historical convention
opposite in sign to each other,  
$$
\mu_S=\mu_s-\mu_q.
$$
Since we use the symbol $S$ also for entropy, henceforth we 
will refrain from mentioning the hyperon number.

\subsection{Resonance decays}
\label{resdec}
In the first instance, we consider hadronic resonances as if they 
were particles with a given well defined mass, {\it e.g.,} their decay width
is insignificant. All hadronic resonances decay rapidly after freeze-out, 
feeding the stable particle abundances.  Moreover,  heavy resonances 
may decay in cascades, where all decays proceed sequentially
from the heaviest to lightest particles. As a consequence, the light
particles obtain contributions from the heavier particles, which have
the form 
\begin{eqnarray}
n_1  &=& b_{2\rightarrow 1} 
 \, ... \, b_{N\rightarrow N-1}
n_N,
\label{n1}
\end{eqnarray}
where $b_{k\rightarrow k-1}$ combines the branching ratio for the $k
\rightarrow k-1$ decay (appearing in \cite{pdg}) with the appropriate
Clebsch--Gordan coefficient.  The latter accounts for the isospin symmetry
in strong decays and allows us to treat separately different charged
states of isospin multiplets of particles such as 
nucleons, Deltas,  pions, kaons, {\it etc}. For
example, different isospin multiplet member states of 
$\Delta$ decay according to the following
pattern:
\begin{eqnarray}
&& \Delta^{++} \rightarrow \pi^+ + p, 
\label{D++} \\
&& \Delta^{+} \rightarrow  {1 \over 3} (\pi^+ + n) + {2 \over 3}
(\pi^0 + p), 
\label{D+} \\
&& \Delta^{0} \rightarrow  {1 \over 3} (\pi^- + p) + {2 \over 3}
(\pi^0 + n), 
\label{D0} \\
&& \Delta^{-} \rightarrow  \pi^- + n.  
\label{D-}
\end{eqnarray}
Here, the branching ratio is 1 but the Clebsch--Gordan coefficients
introduce another factor leading to the effective branching ratios
of 1/3 or 2/3, where appropriate.

A decay such as $\Delta \longrightarrow \pi N$ is easy to deal with, since
only one decay channel is present and the branching ratio is
well known. However, for most of the resonances 
several decay channels appear and the partial widths 
(product of branching ratio with total width) are in general
 less well known, usually  classified as  {\it dominant}, 
{\it large, seen}, or {\it possibly seen,}  \cite{pdg}).
Any statistical hadronization calculation where high-lying 
massive states are relevant should
carefully consider  the dependence of this  uncertainty 
in its results, especially considering
the increasing resonance degeneracy with mass \cite{brongloz}.

Weak hadron decays complicate the physics of hadron yield 
and spectra study  
considerably. The weak interaction life span for strange hadrons
(excluding $K_L$) ranges in  2--10 cm/$c$, which values are 
subject to the time dilation effect in the decaying particle
rest frame. In general these  weak decay feed-down (WDFD) 
occurs both near and far 
enough   from the primary interaction vertex so that
a little known  fraction of decays is counted, or not counted, depending 
on the type of the experiment.
For experiments measuring protons at RHIC, the WDFD  by hyperons is 
large, in the vicinity of 50-60$\%$. For antiprotons at AGS energies the
contamination by antihyperon decay is dominant. 
Similarly, the yield of  pions  includes
a large fraction of  $K_s$ decay products and some hyperon decays. 
Sometimes  experimental results include  corrections 
obtained in simulations of  WDFD,
stripping the reported yields of the WDFD contamination. 
These corrections are   specific to each experiment and even to 
the experimental cuts made in the particular analysis. 
Such corrections introduce  additional 
systematic uncertainties since model dependent assumptions about 
the primary particle spectra must be made. Often enough given these
complications, the reported results  are not corrected for WDFD 
contamination,  leaving us with an interesting 
measurement which is hard to compare to a theoretical model.  

\subsection{Finite resonance width}
\label{widthdis}
For a  particle $i$ which has a finite and significant (thus hadronic) decay 
width $\Gamma_i$, the thermal yield is more appropriately obtained by 
weighting Eq.\,(\ref{nmdef})
over a range of masses in order to take the mass spread into account:
\begin{equation}\label{spread}
\tilde n_i^{\Gamma}=\int\! dM\, 
n(M,g_i;T,\Upsilon_i)\frac{1}{2\pi}\frac{\Gamma_i}{(M-m_i)^2+\Gamma_i^2/4}
\to n_i, \mbox{\ \ for\ \ } \Gamma_i\to 0.
\label{widthn1}
\end{equation}
The exponential thermal weight $n(M,g_i;T,\Upsilon_i)$ 
is  asymmetric around the value of the mass of the particle $M=m_i$ 
and when $\Gamma_i$ and $T$ are of  comparable magnitude
we find that this in  general increases the yield as compared
to the $\Gamma_i\to 0$ limit. But that is not the end of the story.

The    Breit--Wigner  distribution if used with energy independent width means 
that there  would be  a finite probability that the resonance can be formed at 
unrealistically small masses. Since the weight involves a thermal distribution 
$n(M,g_i;T,\Upsilon_i)$ which would contribute in this unphysical domain, one 
has to use, in Eq.\,(\ref{widthn1}), a more physical,  energy dependent width. 
The dominant  energy dependence of the width is due to the decay threshold energy 
phase space factor, dependent  on the angular momentum present  in the decay. 
The explicit form can be seen in the corresponding reverse production 
cross sections \cite{phase1,phase2}. The energy dependent partial width in the 
channel $i\to j$  is to a good approximation: 
\begin{equation}
\label{partwid1}
\Gamma_{i\to j}(M)=b_{i\to j}\Gamma_i
     \left[1-\left(\frac{m_{ij}}{M}\right)^2\right]^{l_{ij}+\frac{1}{2}},
          \quad \mbox{for} \quad M>m_{ij}.
\end{equation}
Here, $m_{ij}$ is the threshold of the decay reaction with branching 
ratio $b_{i\to j}$. For example  for the decay 
of $i:=\Delta^{++}$ into $j:=p+\pi^+$, we have $m_{ij}=m_p+m_{\pi^+}$, while the
branching ratio is unity and the angular momentum released in 
decay is $l_{ij}=1$. From these partial widths 
the total energy dependent width arises,
\begin{equation}
\Gamma_i\to \Gamma_i(M) = \sum_{j}\Gamma_{i\to j}(M).
\end{equation}
For a resonance with width, we thus have replacing Eq.\,(\ref{widthn1}):
\begin{equation}
 n_i^{\Gamma} = \frac{1}{N_i} \sum_{j} 
\int_{m_{ij}}^{\infty} dM \, 
n(M,g_i;T,\Upsilon_i)\frac{\Gamma_{i\to j}(M)}{(M-m_i)^2+[\Gamma_i(M)]^2/4},
\label{widthn}
\end{equation}
and the factor $N$ (replacing $2\pi$) ensures the normalization:
\begin{equation}
\label{norm}
N_i= \sum_{j} 
\int_{m_{ij}}^{\infty}  dM 
\frac{\Gamma_{i\to j}(M)}{(M-m_i)^2+[\Gamma_i(M)]^2/4}.
\end{equation}
The resonance yields with widths in 
general vary considerably  when compared to the yields without widths.
One finds now both enhanced and suppressed yields as compared to the 
limit  $\Gamma_i\to 0$.

\subsection{Chemical  nonequilibrium}\label{alternatives} 
Consider  a large ``pot of boiling quark--gluon soup''. In the
statistical hadronization approach  hadrons `evaporate' with 
an abundance corresponding to the accessible
phase space.   However, the quark equilibrium in the pot in general
precludes that  that just after formation, the 
evaporated hadrons are in chemical yield equilibrium.
In particular, it has long been understood \cite{Koch:1986ud} that the quark
phase space and the hadron phase space are  different.  

It is therefore natural to expect that, if chemical freeze-out is
near a phase transition in which hadrons are formed by quarks recombining according
to hadronic phase space, the hadron gas will be in thermal but not chemical
equilibrium.   Entropy conservation suggests that, if the transition 
is rapid (so the system
can not accommodate {\it e.g.} a   smaller number of available degrees of freedom 
through expansion), particle phase space will be oversaturated significantly above
the point at which detailed balance is achieved \cite{jan_gammaq}.
We could therefore expect $\gamma_{q,s}>1$, with $\gamma_q$ approaching the point
of B-E condensation $(e^{m_{\pi}/2T})$.

Another  physics mechanism  could lead to
 the inapplicability of chemical
equilibrium to strange quark abundance.
Strangeness is produced relatively slowly through kinetic reactions 
(between quarks or hadrons). It is natural to expect the abundance
of strange and antistrange particles to approach 
equilibrium asymptotically from below.
If the system's lifetime is too short,  
strange particle abundance is expected  below
chemical equilibrium, $\gamma_s<1$ \cite{jan_gammas}.    

Strangeness suppression in a small system ({\it e.g.} $pp, pA$), 
as compared to grand canonical yield  Eq.\,(\ref{nmdef}), can 
also be understood within statistical model  as due to phase space
reduction in the canonical statistical model \cite{can1}. 
When strangeness enhancement
in $AA$ collision is reported based on comparison 
with respect to these smaller systems, an artificial 
enhancement is expected  \cite{can2}. 
This artificial enhancement effect depends
in a specific fashion on volume, 
particle strangeness, and reaction 
energy. This mechanism does not  describe
the  $AA$ data in detail \cite{jancan}. Recent experimental results 
confirm this. Both the predicted rapid approach to the grand canonical limit
 as function of the  number of
participants, and the   enhancement increase of {\it e.g.} the 
$\Xi, \overline{\Xi}$ with decreasing energy  disagree decisively  with     
  experimental  40 GeV and 158 GeV SPS  results  \cite{spscan1,spscan2,spscan3}. With 
these experimental results  the effect of strangeness
enhancement in $AA$ reactions is confirmed to be the  result
 of kinetic production processes.

\section{Statistical hadronization model}\label{model}
\subsection{Model-data comparisons}

In the most general version of the statistical model, 
(excluding in this consideration  
charmed quarks), we have six independent parameters 
which have to be determined by experimental
data: $T,\lambda_{q,s},\gamma_{q,s},\lambda_{I3},$ 
and an overall volume parameter normalizing particle 
yields. This  volume $V$ is proportional to the magnitude of
the transverse geometric  area of the reacting system
determined by the size of the nuclei and the event trigger, 
and it depends on the model dependent 
longitudinal  rapidity acceptance relation with the longitudinal 
volume extend. This  dependence of  $V$  on  non-trivial 
and also experiment-specific factors makes it tempting to eliminate 
it in a  fit of  particle ratios.
For  yield studies  which include the full particle multiplicity,
the absolute yields can, however, be fitted and the volume parameter is determined 
by the trigger condition which fixes the number of participants.

One of the remaining 5 parameters, $I_3$,  is fixed by the proton 
to neutron asymmetry in  the reaction.
 No matter how uncertain we are about the longitudinal dynamics,  the ratio 
$(\langle d\rangle -\langle\bar d\rangle)/(\langle u\rangle -\langle\bar u\rangle)$
(where $\langle d\rangle$ refers to the number of valance $d$-quarks, etc.)
is fixed and we accommodate this by fixing the charge to baryon ratio 
of the colliding nuclei. Another constraint is strangeness conservation, that is 
$\langle s\rangle =\langle\bar s\rangle$. When implementing this condition one
generally fixes the value of $\lambda_s$. Here there are, however, some caveat.

First, since strangeness decays weakly, the final strangeness yield is, at finite
baryon number, asymmetric. Considering that there are more   strange
 baryon weak decays than strange antibaryon weak decays, the true weak decay correction 
situation is that the final state has $\langle s\rangle <\langle\bar s\rangle$. 
In consequence of experimental corrections which tend 
to counterbalance this effect  one should perhaps not impose exact
strangeness conservation but allow a measure of uncertainty when implementing
strangeness conservation. Secondly, because of 
baryon asymmetry, the rapidity distribution of strange and antistrange baryons 
needs not to be exactly identical. Given that the baryon number at RHIC is 
expected to be mostly in fragmentation regions, this  is where one could 
expect some excess of $\langle s\rangle > \langle\bar s\rangle$.

We obtain  the fit parameters from the experimental 
  by minimizing  $\chi^2$:
When $f_i (T,\lambda,\gamma)$ is the model prediction 
for particle $i$, $n_i$ is the experimental value,
and $\Delta_{stat,sys}$ are the statistical and systematic errors, 
\begin{equation}
\chi^2 \equiv \sum_i \left(  \frac{f_i(T,\lambda,\gamma)-n_i}
                {  \Delta_{stat}+\Delta_{sys}} \right)^2
\end{equation}
The best fit parameters $T,\lambda_i,\gamma_i$ are those 
which minimize $\chi^2$ while respecting constraints described
above.

\begin{figure}[t]
\begin{center}
\vskip 0.3cm
\hspace*{-.2cm}\epsfig{width=11cm,figure=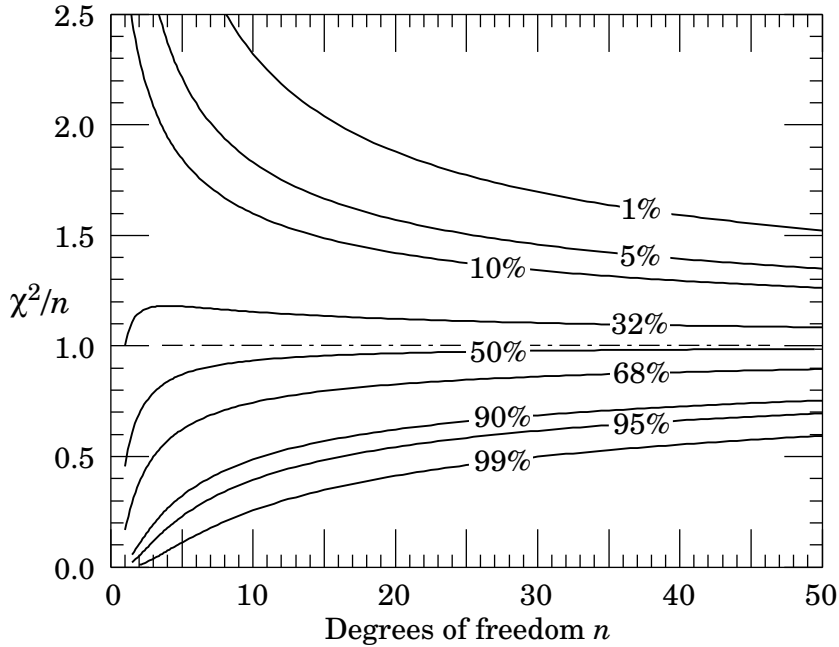}
\caption{\label{statsig} 
Lines of constant statistical significance $P$ [\%] as a function of the number of 
degrees of freedom $n$ and $\chi^2/n$ \protect\cite{pdg}.
}
\end{center}
\end{figure}

The previous section makes it clear that any fit of statistical
 model parameters to data should, in principle, 
allow for the possibility that some, or all 
quantum numbers are not in chemical equilibrium.
In other words, equilibrium should be an eventual 
result of the study, rather than it's assumption.
However,  fixing the non equilibrium parameters to 
a tacit value such as $\gamma_{q,s}=1$ reduces 
 seemingly (that is by assumption)
the number of parameters, which increases  $n$, 
the number of degrees of freedom (d.o.f)  in the fit.
To compare  fits which use $\gamma_{q,s}$ with those which fix it in case
that  $n$ is not very large, it is necessary to consider a comparison standard 
independent of the value of $n$.
Such a standard is provided by statistical significance $P$ shown 
in Fig. \ref{statsig}, Ref. \cite{pdg} . Each line is drawn for a
fixed value of $P$ [\%]  defined as the likelihood, 
given the validity of the model, and a purely
 Gaussian error source, of finding the corresponding 
 $\chi^2/n$, or a smaller value. 
 
As can be seen $P$  reaches an asymptotic 
value for $n\to \infty$, where it rapidly approaches  100\%  
for  $\chi^2/n \leq 1$ and rapidly approaches 0 otherwise.
However, the typical number of degrees of freedom in statistical
 model fits is considerably below the asymptotic limit, making the 
statistical significance dependent on $n$. For this reason 
 perfectly acceptably-looking graphs can have 
 an unacceptably small statistical
significance. For example, for a fit  with {\it e.g.}   5
 measurements and 3 parameters to have $P=90\% $ it must come up with  
$\chi^2/n \leq 0.1$, that is the fit 
 must hit the center of each measurement to 
be considered based on a good physical model. The 
range $\chi^2/n \sim 1$ corresponds in this case to $P=30\%$,
which means that most likely the physical model used is false
or the data inaccurate or too precisely stated.

Statistical significance, per se, is not a ``proof of validity'' of a model.   
 It is, however, a discriminating force.  If a fit has a `small' statistical
 significance, well below 60\%, the fit's validity and/or experimental 
data should be  questioned, {\it e.g.} are there unaccounted for 
systematic errors?   Are particular data points spoiling the fit, 
and if so, why? Thus study of $P$ must supersede all  ``looks nice''
arguments.   And if a markedly higher statistical significance 
can be achieved by varying from a tacit value  a physically motivated fit
parameter, this can be taken as evidence that the physical scenario 
underlying this parameter is relevant.

\subsection{Sample of hadron yield fit results}\label{fits}
The table \ref{results_table130} and
 Fig. \ref{results_table130} were also presented in  \cite{bari}. 
Similar results are found in fits
to SPS data \cite{observing}.
We note that Table \ref{results_table130} shows that 
strangeness under-saturation models appear to be
 ruled out   at RHIC-130. 
The fitted parameter $\gamma_s$ for a non equilibrium fit
is well above unity, and it is  slightly 
above above unity  in fits where $\gamma_q=1$.
This latter observation  is corroborated by another
 fit of RHIC data  \cite{kaneta}.
We recall that SPS and AGS results are best fitted with $\gamma_s/\gamma_q<1$
\cite{Bec04}. We see in table \ref{results_table130}  
that chemical equilibrium model fits yield   relatively 
low statistical significance $P$, which is
increasing as we step up the non equilibrium,
 introducing first $\gamma_{ s}\ne 1$ 
and than  $\gamma_{q}\ne 1$ .
 
Even though there are some serious
differences in confidence level, the three models 
 look nearly indistinguishable  in Fig. \ref{results_table130}.
  Fig. \ref{statsig} shows that at $\sim 10$ degrees of freedom, one needs
a $\chi^2/n$ DoF considerably below 1, which requires that  
the average data point appears
 within the error bar.     
\begin{table*}[bt]
\caption{
\label{results_table130} RHIC-130 GeV hadronization parameters}
\vspace*{0.2cm}\small
\begin{tabular}{|c| c c| c c| c c|}
\hline
 &        \multicolumn{2}{c}{$\gamma_{q,s}$ vary }  &\multicolumn{2}{c}{$\gamma_{q,s}=1$ }  & \multicolumn{2}{c|}{$\gamma_{q}=1$, $\gamma_s$ varies }  \\
parameter & $\Gamma=0$ & $\Gamma$  \cite{pdg} & $\Gamma=0$ & $\Gamma\ne 0$  & $\Gamma=0$ & $\Gamma\ne 0$  \\
\hline
T [MeV]         & 133 $\pm$ 10 &  135 $\pm$ 12 & 158 $\pm$ 13 & 0.157 $\pm$ 15 & 152 $\pm$ 16 & 153 $\pm$ 23 \\
$10^4 (\lambda_{q}-1)$     & 708 $\pm$ 342 & 703 $\pm$  337  &   735 $\pm$ 390 & 730 $\pm$  382 & 724 $\pm$  373 &  721 $\pm$     363\\
$10^2(\lambda_s$\footnote{Found by solving for strangeness}$-1)$
                &    3.132973   & 3.203555 &  2.636207 &2.788897 &2.95346 &  3.00848  \\
$\gamma_{q}$    & 1.66 $\pm$ 0.013 &  1.65 $\pm$ 0.030  & 1 & 1 & 1 & 1 \\
$\gamma_{s}$      & 2.41 $\pm$ 0.61 &   2.28 $\pm$  0.46  & 1 & 1 & 1.17 $\pm$ 0.30 &1.10  $\pm$ 0.25\\ 
$10^4 (\lambda_{I_3}-1)$ &  30 $\pm$ 305 & 28 $\pm$  293 & 59 $\pm$  564 & 53 $\pm$  508 &  64  $\pm$  525 &  59  $\pm$    481 \\
\hline
 & \multicolumn{6}{c|}{fit relevance}\\ 
\hline 
$N-p=\mbox{n}$                                               & 16-5 &    16-5   & 16-3   & 16-3   & 16-4 & 16-4 \\   
 $\chi^2/$n             &  0.4243  & 0.4554 &  1.0255   & 0.8832 &  0.6067 &  0.7301 \\ 
 $P$[\%]   & 94.61   & 93.07 &  42.25  &  57.05 &  83.85 &  72.32 \\
\hline
\end{tabular}
 \end{table*}

\begin{figure}[t]
\begin{center}
\vskip 0.3cm
\psfig{width=11cm,figure=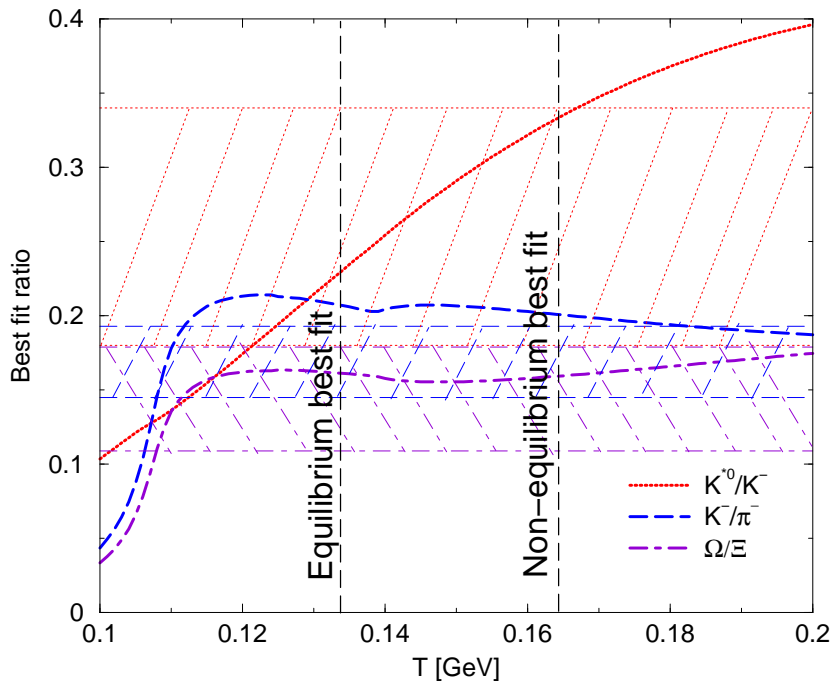}
\caption{\label{pdat_T} Sensitivity of a measurement of
$K^{*0}/K^-,K^-/\pi^-,\Omega/\Xi^-$  on Temperature $T$,
other statistical hadronization model parameters have been
optimized to best fit the other experimental RHIC-130 
data. 
}
\end{center}
\end{figure}
\begin{figure}[t]
\begin{center}
\vskip 0.3cm
\psfig{width=11cm,figure=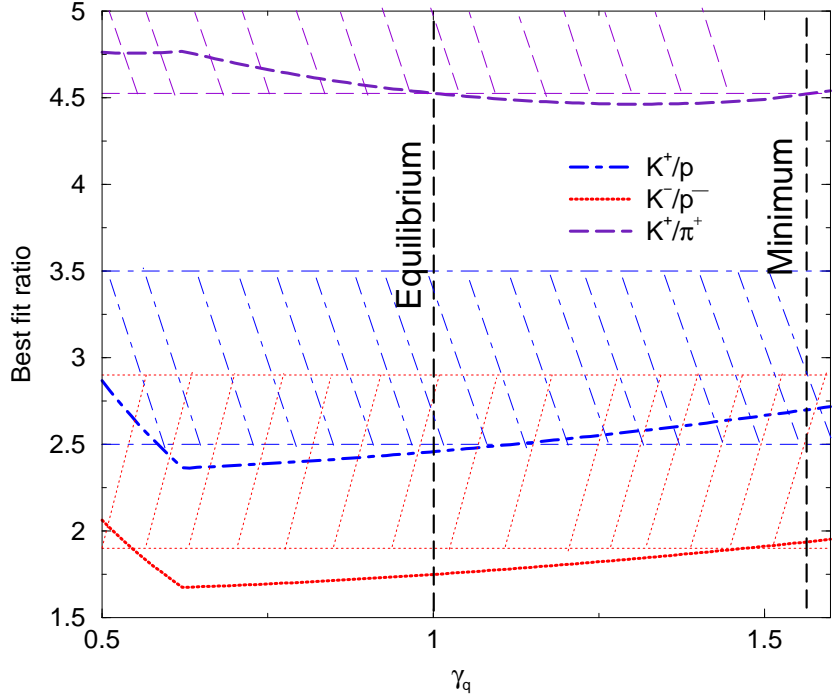}
\caption{\label{pdat_gammaq}Sensitivity of a measurement of
$K^{+}/p,K^-/\bar p^-,K^+/\pi^+$  on the value of $\gamma_q$,
the chemical light quark pair occupancy parameter,
other statistical hadronization model parameters have been
optimized to best fit the other experimental RHIC-130 
data.   
}
\end{center}
\end{figure}
\begin{figure}[t]
\begin{center}
\vskip 0.3cm
\psfig{width=11cm,figure=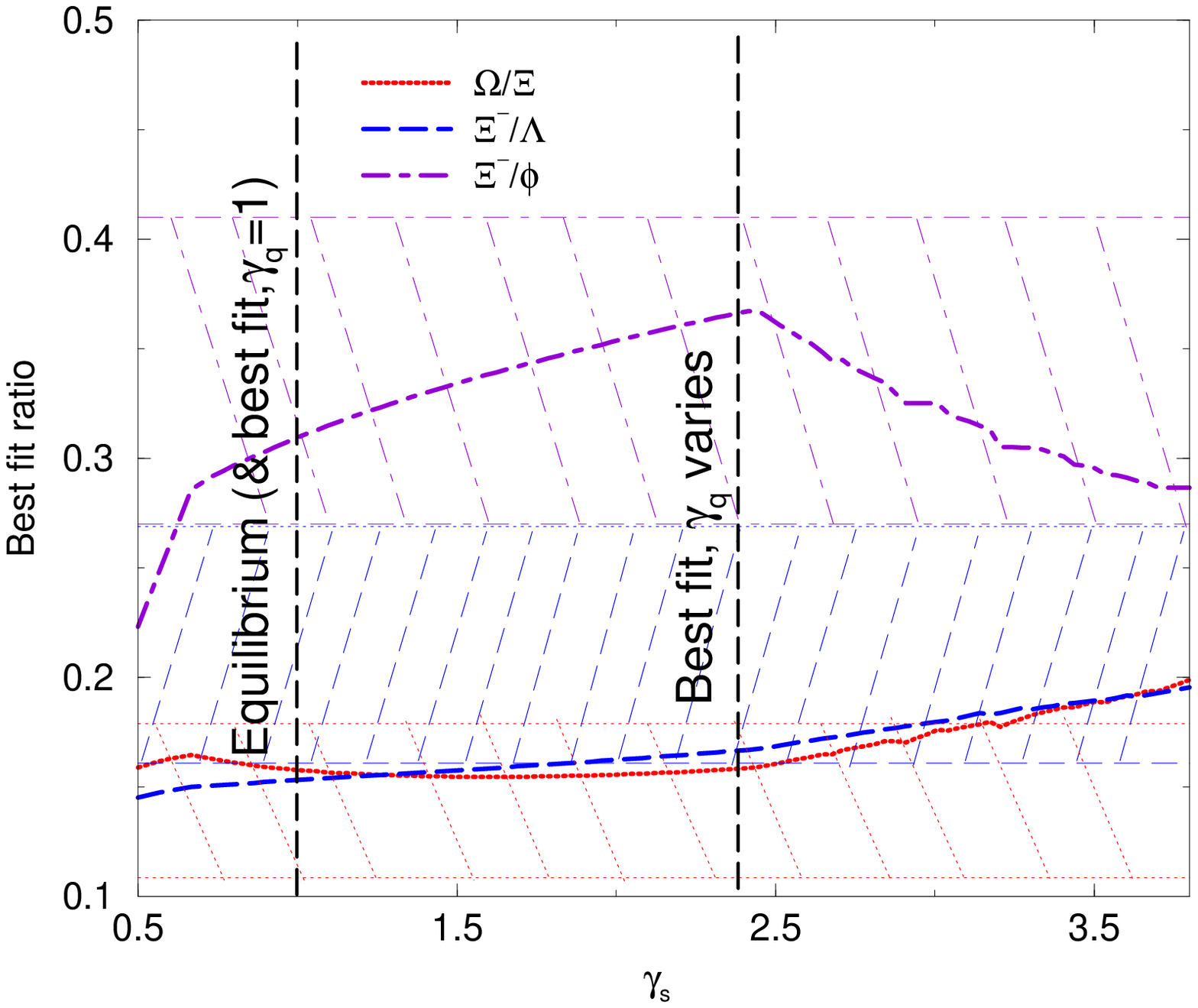}
\caption{\label{pdat_gammas}  Sensitivity of a measurement of
$\Omega/\Xi^-,\Xi^-/\Lambda,\Xi^-/\phi $  on the value of  $\gamma_s$,
the chemical strange quark pair occupancy parameter,
other statistical hadronization model parameters have been
optimized to best fit the other experimental RHIC-130 
data.  
}
\end{center}
\end{figure}

Given the present data quality,  it is not
 possible to determine with precision values of  $\gamma_{q,s}$ , and hence
to differentiate full non-equilibrium from strangeness non-equilibrium.
We can try constraining the
 fit parameters further by examining sensitivity 
of data point to them in detail.
Figures \ref{pdat_T}, \ref{pdat_gammaq} and \ref{pdat_gammas} aim to to  that.
To prepare them, we have performed the best fit  
at a given fixed  $T$ for Fig. \ref{pdat_T}, $\gamma_q$ for \ref{pdat_gammaq} 
and $\gamma_s$ for \ref{pdat_gammas}, varying all the other parameters.  
The model  data points are than graphed against the parameter in question 
to show sensitivity to the parameter in this observable.
\begin{figure}[t]
\begin{center}
\vskip 0.3cm
\hspace*{-.2cm}\psfig{width=11cm,figure= 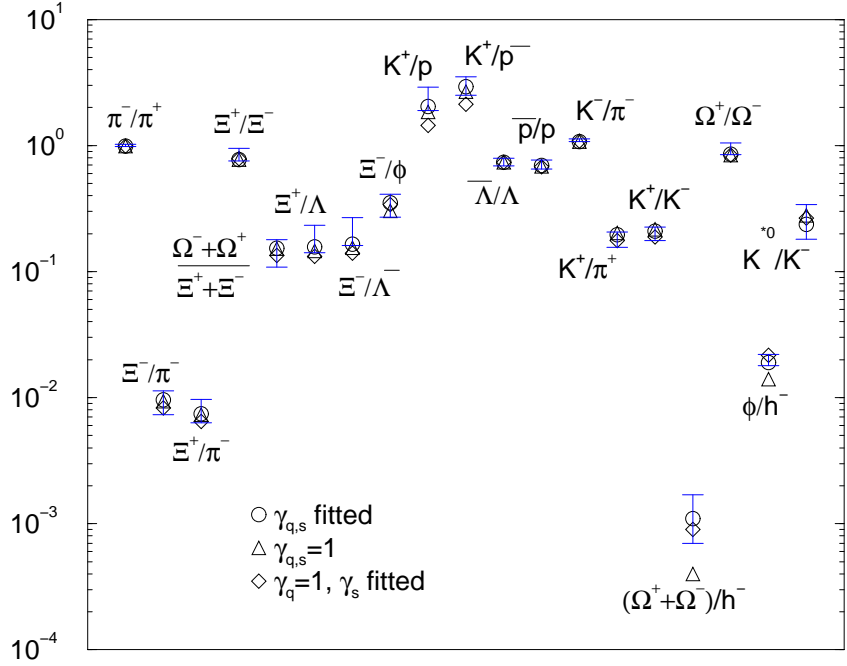}
\caption{\label{plot130} A comparison of 130 GeV Au-Au 
data and the three different statistical models -- all
results `look good', even if statistical significance 
varies considerably, see bottom of table
   \protect\ref{results_table130}.
}
\end{center}
\end{figure}
It is immediately apparent that, in models with chemical 
non-equilibrium, parameter correlation makes it possible
to model some particle ratios over a wide range of any parameter.   
 For instance, Fig. \ref{pdat_T} shows that
once temperature reaches $\sim 130$ MeV, it is possible to model 
ratios such as $K/\pi$ and $\Xi/\Omega$ using any
temperature within the range.    It is only when several
 similar ratios are fitted together that parameters
become disentangled.

Figs \ref{pdat_T},\ref{pdat_gammaq},\ref{pdat_gammas} also show that 
some data points are better at disentangling fit parameters than others.
Fig. \ref{pdat_T} shows that resonances (here $K^*$)  are ideal for  constraining 
the freeze-out temperature.
If compared to a stable particle whose quark composition is identical,
 all chemical factors ($\lambda,\gamma$) cancel out independently
of equilibrium, so the ratio is a function of temperature and flavor content
alone.   When the $K^*/K$ ratio has been measured to a greater precision, 
it will be possible to definitely constrain the freeze-out temperature.   
This could be sufficient to distinguish an equilibrium scenario from one
 where $\gamma_q$ is needed to explain hadronic abundance.

Constraining chemical potentials, in particular to ascertain directly 
if $\gamma_{q,s}\ne 1$ are present, is more difficult, since these parameters are
strongly correlated with temperature. Thus measurement of resonances also fixes
this issue, in an indirect way. In addition, 
as Fig. \ref{pdat_gammas} shows, $\gamma_s$ can be constrained by 
measuring ratios of particles of similar mass, but different 
strangeness content, such as $\Xi$ and $\Lambda$,
while $\gamma_q$ (Fig. \ref{pdat_gammaq}) is sensitive to 
ratios between baryons and mesons.   At the moment, the error
 bars for these data points are too large to allow
disentangling $T$, $\gamma_q$ and $\gamma_s$.

\section{How to SHARE}\label{SHARE} 
In the following years, the  high-statistics RHIC date, 
beginning with 200 GeV run  will be thoroughly analyzed, 
leading to better determination  of the yields
of many stable particles and resonances.   
As shown in the previous section, these measurements 
should be sufficient to distinguish between 
models described above.  
In addition,  SPS and eventually the future GSI facility, 
as well as higher energy LHC measurements promise an extensive energy scan, 
allowing a full assessment of the applicability of different 
statistical models in a wide range of $\sqrt{s}$, as well as the detection of any
systematics in the fit parameters.

We have argued that, if the purpose of statistical analysis is to observe
an eventual phase transition, any assumptions about chemical equilibration
have to be tested against experimental data.  
We have also argued that statistical hadronization calculations
require a very detailed input of the hadronic spectrum, in particular in
regard to the uncertainty in degeneracy and decay patter of the higher-lying states
\cite{pdg,brongloz} as well as experiment-specific weak-decay feed-downs.

To get the correct physics out the experimental data, and to
 ascertain the physical significance behind the statistical
model, these approaches need to be compared with experimental 
data using a standard procedure. We have proposed an open source 
general model for:  Statistical HAdronization with REsonances (SHARE) \cite{share}.
This statistical hadronization
package is suitable for comparative analysis of heavy ion collisions in a variety of
systems, energy ranges, and equilibrium assumptions. 
A cross check with SHARE requires that:
\begin{enumerate}
\item the resonance decay tree   be the same for the different 
models under consideration, so as all systematic effects in the fit
parameters and the quality of the fit deriving from choice of 
resonance decay are under control,
\item experiment-specific weak decay acceptances should
 be treated separately for each experimental data-point, to minimize 
systematic errors on fit parameters deriving from
experimental effects,
\item each model's ability to fit data should be compared 
in a way that takes into account the different number of
 degrees of freedom within each model.
Statistical significance is a comparison criterion 
which satisfies this requirement,
\item the fitting program should be able to 
fit each model's fit parameter, or constrain it 
using conservation laws (if applicable),
\item the fitting program should be able to evaluate
the fit's sensitivity to each parameter checking 
for  false minima, etc. and  
employing  $\chi^2$ profiles and contours,
\item the fitting program should be able to 
test the fit's sensitivity to each data point.  
This is invaluable both to determine the consistency
 of the fit, and to detect systematic violations 
indicating possible novel microscopic formation mechanisms,
\item to better assess the physical soundness of the statistical
 picture, bulk quantities (total number 
of negative/positive/neutral particles, entropy, pressure,
energy density etc.should also be calculated.
\end{enumerate}
All these features are part of the package   SHARE \cite{share}.
Moreover we offer three different evaluation of particle multiplicities
(SHARE FORTRAN, Mathematica SHARE, and web based SHARE - derived from 
the FORTRAN version of SHARE).  SHARE   calculates yields, 
ratios and bulk quantities in terms of thermodynamic parameters, 
Fortran version performs fits, and explores parameter sensitivity to data
($\chi^2$ profiles, contours etc.). The  resonance decay trees, and 
weak corrections are allowed form and Fortran version requires in the fit
the  experimental ratios as input.
 
The reader interested in this field is encouraged to use SHARE in study 
by means of   statistical hadronization models  new experimental data.
The program is available online, at\\
\texttt{http://www.physics.arizona.edu/$\sim$torrieri/SHARE/share.html}\\
\texttt{http://www.ifj.edu.pl/Dept4/share.html}\\
We hope that this effort will contribute in achieving a better 
understanding of the applicability and physical significance of
 the statistical particle production picture. 
\subsection*{Acknowledgments}
Work supported in part by grants from: the U.S. Department of
Energy  DE-FG03-95ER40937 and DE-FG02-04ER41318,  NATO Science Program PST.CLG.979634.  LPTHE, Univ.\,Paris 6 et 7
is: Unit\'e mixte de Recherche du CNRS, UMR7589. 
G. Torrieri thanks the organizers of the Krakow school 
for theoretical physics for the opportunity
to present this work, and  Charles Gale and SangYong Jeon for helpful 
and stimulating discussions. We thank Wojtek Broniowski and Wojtek 
Florkowski for a  fruitful collaboration on the SHARE project.


\begin{thebibliography}{10}


\bibitem{KMR03}
J. Kapusta, B. Muller and J. Rafelski, 
{\it Quark--Gluon Plasma: Theoretical Foundations}
An annotated reprint collection, Elsevier, 
ISBN: 0-444-51110-5, 836 pages, 2003.
\bibitem{wbwfhag}
W. Broniowski and W. Florkowski, Phys. Lett. {\bf B490}, 223 (2000).

\bibitem{Letessier:gp}
J.~Letessier and J.~Rafelski,
Cambridge Monogr.\ Part.\ Phys.\ Nucl.\ Phys.\ Cosmol.\  {\bf 18},  1 (2002).

\bibitem{hagedorn}
R.~Hagedorn,
Nuovo Cim.\ Suppl.\  {\bf 3}, 147 (1965).

\bibitem{schne} E. Schnedermann, J. Sollfrank, and U. Heinz,
Phys. Rev. C  {\bf 48}, 2462 (1993).

\bibitem{pbmags} 
P. Braun-Munzinger, J. Stachel, J. P. Wessels, and N. Xu, 
Phys. Lett. B {\bf 344}, 43 (1995); Phys. Lett. B {\bf 365}, 1 (1996).

\bibitem{rafacta27}
J.~Rafelski, J.~Letessier, and A. Tounsi, 
Acta Phys. Pol. B {\bf 27}, 1035 (1996).


\bibitem{BL} T. Cs\"{o}rg\H{o} and B. L\"{o}rstad, 
Phys. Rev. C {\bf 54}, 1390 (1996).

\bibitem{cest} 
J. Cleymans, D. Elliott, H. Satz, and R. L. Thews, Z. Phys. 
{\bf 74}, 319 (1997).

\bibitem{rafacta28} 
J. Rafelski, J. Letessier, and A. Tounsi, 
Acta Phys. Pol. B {\bf 28}, 2841 (1997).

\bibitem{cr} 
J. Cleymans and K. Redlich,
Phys. Rev. Lett. {\bf 81}, 5284 (1998).

\bibitem{PBM99}
P. Braun-Munzinger, I. Heppe and J. Stachel,  Phys. Lett. B
\textbf{465}, 15 (1999).

\bibitem{yg} 
G. D. Yen and M.~Gorenstein, 
Phys. Rev. C {\bf 59}, 2788 (1999).

\bibitem{gazacta}
M. Ga\'zdzicki and M.~Gorenstein, 
Acta Phys. Pol. B {\bf 30}, 2705 (1999). 

\bibitem{gaznpa} 
M. Ga\'zdzicki, 
Nucl. Phys. A {\bf 681}, 153 (2001).

\bibitem{finland1}  
F. Becattini, J. Cleymans, A. Keranen, E. Suhonen, and K. Redlich, 
Phys. Rev. C {\bf 64}, 024901 (2001).

\bibitem{finland2}  
F. Becattini, J. Cleymans, A. Keranen, E. Suhonen, and K. Redlich, 
arXiv:hep-ph/0011322.

\bibitem{wfwbmm} W. Florkowski, W. Broniowski, and M. Michalec, 
Acta Phys. Pol. B {\bf 33}, 761 (2002).

\bibitem{csorgo} T. Cs\"{o}rg\H{o}, 
Heavy Ion Phys. {\bf 15}, 1 (2002).

\bibitem{zakopane}
W.~Broniowski, A.~Baran, and W.~Florkowski, Acta Phys. Pol. B 
{\bf 33}, 4235 (2002).

\bibitem{cool}
J.~Rafelski, J.~Letessier,  and A.~Tounsi,
Acta.\ Phys.\ Pol.\ A {\bf 85}, 699  (1994).
J.~Letessier, J.~Rafelski and A.~Tounsi,
Phys.\ Rev.\ C {\bf 50}, 406 (1994);
[arXiv:hep-ph/9711346].

\bibitem{PBMRHIC}
D.~Magestro,
J.\ Phys.\ G {\bf 28}, 1745 (2002).

\bibitem{wbwf}
W.~Broniowski and W.~Florkowski,
Phys.\ Rev.\ Lett.\  {\bf 87}, 272302 (2001).

\bibitem{pdg}
K.~Hagiwara {\it et al}.,  Particle Data Group Collaboration,
Phys.\ Rev.\ D {\bf 66}, 010001 (2002), 
see also earlier versions, note that the MC identification scheme for
most hadrons was last presented in 1996.

\bibitem{brongloz}
W.~Broniowski, W.~Florkowski and L.~Y.~Glozman,
arXiv:hep-ph/0407290.


\bibitem{phase1}
H.~Terazawa,
Phys.\ Rev.\ D {\bf 51}, 954 (1995).

\bibitem{phase2}
G.~J.~Gounaris and J.~J.~Sakurai,
Phys.\ Rev.\ Lett.\  {\bf 21}, 244 (1968).

\bibitem{Koch:1986ud}
P.~Koch, B.~Muller and J.~Rafelski,
Phys.\ Rept.\  {\bf 142}, 167 (1986).

\bibitem{jan_gammaq}
Phys.\ Lett.\ B {\bf 475}, 213 (2000)
[arXiv:nucl-th/9911043].

\bibitem{jan_gammas}
J.~Rafelski,
Phys.\ Lett.\ B {\bf 262}, 333 (1991);\\
J.~Letessier, A.~Tounsi, U.~W.~Heinz, J.~Sollfrank and J.~Rafelski,
Phys.\ Rev.\ D {\bf 51}, 3408 (1995)
[arXiv:hep-ph/9212210];\\
F.~Becattini, M.~Gazdzicki and J.~Sollfrank,
Eur.\ Phys.\ J.\ C {\bf 5}, 143 (1998)
[arXiv:hep-ph/9710529];\\
J.~Cleymans, B.~Kaempfer, P.~Steinberg and S.~Wheaton,
J.\ Phys.\ G {\bf 30}, S595 (2004)
[arXiv:hep-ph/0311020].

\bibitem{can1}
J.~Rafelski and M.~Danos,
Phys.\ Lett.\ B {\bf 97}, 279 (1980).

\bibitem{can2}
S.~Hamieh, K.~Redlich and A.~Tounsi,
Phys.\ Lett.\ B {\bf 486}, 61 (2000)
[arXiv:hep-ph/0006024].

\bibitem{can3}
K.~Redlich and A.~Tounsi,
Eur.\ Phys.\ J.\ C {\bf 24}, 589 (2002)
[arXiv:hep-ph/0111261].

\bibitem{jancan}
J.~Rafelski and J.~Letessier,
J.\ Phys.\ G {\bf 28}, 1819 (2002)
[arXiv:hep-ph/0112151].

\bibitem{spscan1}
F.~Antinori {\it et al.}  [NA57 Collaboration],
Phys.\ Lett.\ B {\bf 595}, 68 (2004)
[arXiv:nucl-ex/0403022].

\bibitem{spscan2}
G.~E.~Bruno  [NA57 Collaboration],
J.\ Phys.\ G {\bf 30}, S717 (2004)
[arXiv:nucl-ex/0403036];

\bibitem{spscan3}
T.~Virgili {\it et al.}  [NA57 Collaboration],
``Recent results from NA57 on strangeness production in p A and Pb Pb
collisions at 40-A-GeV/c and 158-A-GeV/c,''
arXiv:hep-ex/0405052;\\
D.~Elia {\it et al.}  [NA57 Collaboration],
``Strange particle production in 158 and 40 $A$ GeV/$c$ Pb-Pb and p-Be
collisions,''
Contribution to the proceedings of the "Hot Quarks 2004" Conference, July 18-24 2004, New Mexico, USA, submitted to Journal of Physics G,
arXiv:nucl-ex/0410034.

\bibitem{bari}
G.~Torrieri and J.~Rafelski,
arXiv:hep-ph/0409160.

\bibitem{observing}
J.~Letessier and J.~Rafelski,
Int.\ J.\ Mod.\ Phys.\ E {\bf 9}, 107 (2000)
[arXiv:nucl-th/0003014].


\bibitem{kaneta}
M.~Kaneta and N.~Xu,
arXiv:nucl-th/0405068.

\bibitem{Bec04}
F.~Becattini, M.~Gazdzicki, A.~Keranen, J.~Manninen and R.~Stock,
 Phys.\ Rev.\ C {\bf 69}, 024905 (2004)
[arXiv:hep-ph/0310049].
 

\bibitem{share}
G.~Torrieri, W.~Broniowski, W.~Florkowski, J.~Letessier and J.~Rafelski,
arXiv:nucl-th/0404083.


\end{thebibliography}
\end{document}